\documentclass[sn-mathphys,authoryear]{sn-jnl}% Math and Physical Sciences Reference Style

\usepackage{epsfig}
\usepackage{bm}
\usepackage{amsmath,amssymb,amsfonts}
\usepackage{xcolor}
\usepackage{manyfoot}
\usepackage{graphicx}
\usepackage{amsthm}
\usepackage{textcomp}

\begin{document}

\title{Cross-helicity and chaotic dynamics of full-disc solar magnetic field}

\author*{\fnm{Alexander} \sur{Bershadskii}}

\affil{\orgname{ICAR}, \orgaddress{\city{P.O. Box 31155, Jerusalem}, \postcode{91000}, \country{Israel}}}

\abstract{
Using the results of laboratory experiments and direct numerical simulations, as well as observations of the full-disc solar magnetic field and sunspot number dynamics, it is demonstrated that cross-helicity can dominate the decaying part of the frequency power spectra of the magnetic field generated by a magnetohydrodynamic (MHD) dynamo in chaotic/turbulent swirling flows for sufficiently strong MHD turbulence (including the solar dynamo). The theoretical consideration is based on a Kolmogorov-like phenomenology within the framework of the distributed chaos concept.  It is also shown that the solar full-disc magnetic field for the last two solar cycles with weak magnetic activity exhibits deterministic chaotic behavior concentrated around the equator. }

\maketitle

\section{Introduction}

   There is a consensus that a magnetohydrodynamic (MHD) dynamo generates the Sun's global magnetic field (see for an excellent recent review \citealp{cs}). While the global magnetic dynamo is likely situated in the convection zone (or just beneath it), there are also local near-surface MHD dynamos associated with the active regions \citep{b1}. Naturally, a fundamental question regarding the predictability of solar activity is also based on the dynamo models (see, for instance, a very recent paper by \citealp{obr2} and references therein). \\

    The primary source of an MHD dynamo is the transformation of the kinetic energy of the electrically conductive fluid (plasma) into magnetic energy. Therefore, it can be expected that the (averaged) cross-helicity density
\begin{equation} 
 h_{cr} =\langle {\bf u} \cdot {\bf b} \rangle, 
 \end{equation}
which relates the velocity ${\bf u}$ and magnetic ${\bf b}$ fields (and is a fundamental invariant of ideal magnetohydrodynamics), should play a significant role in this process.  However, the significance of cross-helicity for the MHD dynamo is often underestimated (see, for instance, a recent review \citealp{yok} and references therein).\\

  For the astrophysical applications (in the Sun's and stars' interiors, for instance), the global rotation, which violates the cross-helicity invariance, could be the main reason for ignoring the contribution of cross-helicity to MHD dynamo processes. In this case, however, a generalized cross-helicity
 \begin{equation} 
 \tilde{h}_{cr} =\frac{1}{2}\langle {\bf u}\cdot {\bf b} \rangle + \boldsymbol\Omega \cdot {\bf A}(t) 
\end{equation} 
 can be introduced \citep{shebalin}. Here, $\boldsymbol\Omega$ is the constant (global) angular velocity, and ${\bf A}(t)$ is a spatially uniform yet time-dependent part of the magnetic vector potential. This generalized helicity is still an invariant of ideal MHD in a rotating frame of reference \citep{shebalin}.\\ 
 
  Another reason may be the close relationship between the very existence of {\it nonzero} average cross-helicity and the violation of the global (net) reflectional symmetry of the system (the same problem exists with another ideal MHD invariant -- magnetic helicity \citealp{b1}). This problem often arises in numerical simulations. Since the breaking of the {\it local} reflectional symmetry is an inherent property of the chaotic/turbulent flows, the problem can be solved just for these flows (Section 2.2 of the present paper). \\
  
   It is known that the MHD dynamo is excited due to the nonlinear instabilities and is developed through the deterministic chaos states (see, for instance, Ref. \citealp{yvw,cs} and references therein). For bounded and smooth dynamical systems, one of the simplest ways to identify the existence of deterministic chaos is by calculating their power spectra. 
   
   The continuous exponential frequency spectrum 
\begin{equation}
E(f) \propto \exp-(f/f_c)   
\end{equation} 
is a good indication in this case \citep{oh,mm1,mm2,mm3}.\\

   Let us consider, as an example, the simple Lorenz model  \citep{lor}
\begin{equation}
\frac{dx}{dt} = \sigma (y - x),~~      
\frac{dy}{dt} = r x - y - x z, ~~
\frac{dz}{dt} = x y - b z    
\end{equation}  

 %%%%%%%%%%%%%%% 1 %%%%%%%%%%%%%%%%%%
\begin{figure} \vspace{-2cm}\centering \hspace{-0.7cm}
\epsfig{width=.8\textwidth,file=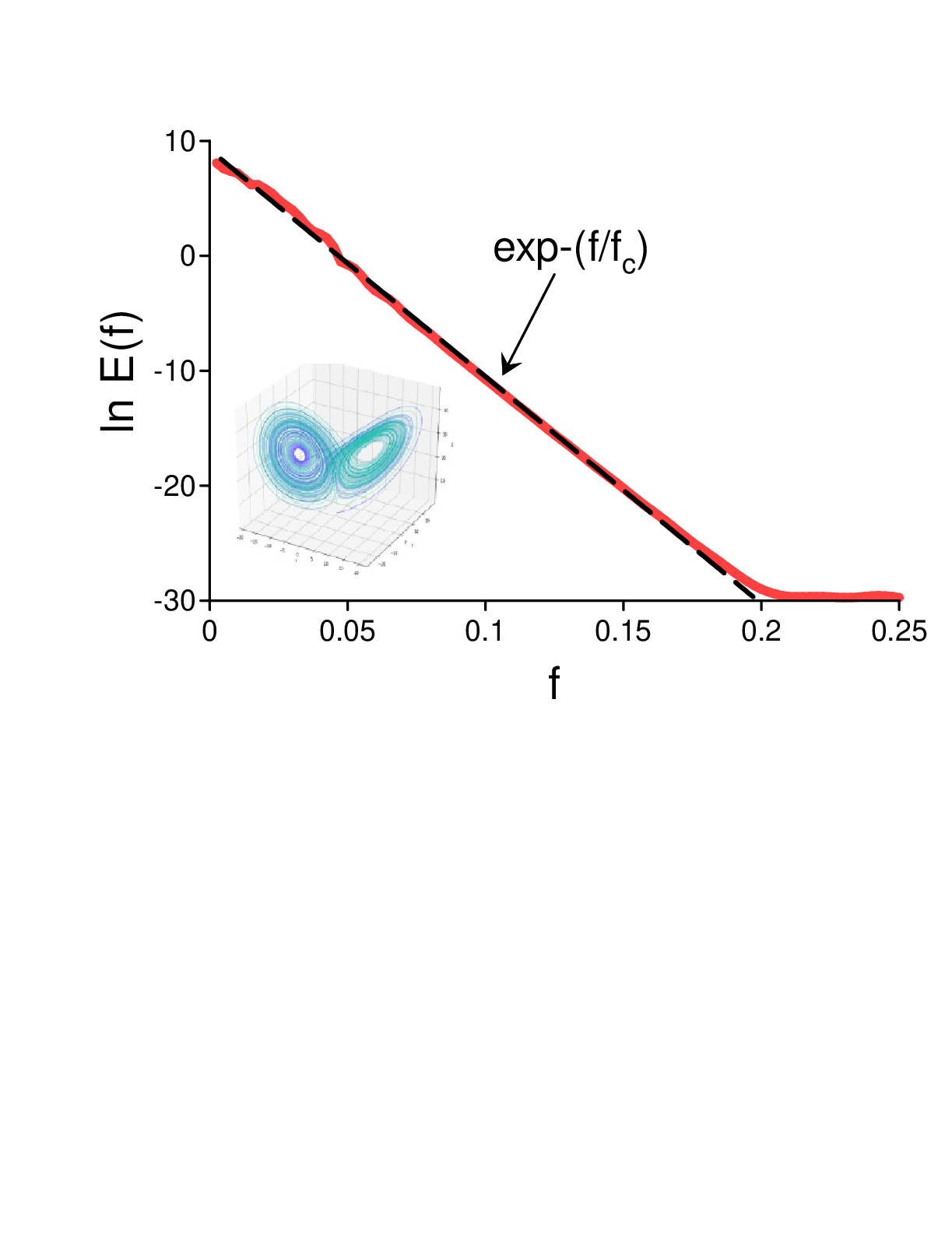} \vspace{-7.2cm}
\caption{Power spectrum of the $x(t)$ variable for the Lorenz system Eq. (4) in a chaotic state. The inset shows the corresponding phase portrait of the system.} 
\end{figure}
%%%%%%%%%%%%%%%%%%%%%%%%%%%%%%%%%%% 

    Figure 1 shows (on semi-logarithmic axes) the power spectrum of the variable $x(t)$ for the Lorenz system in a chaotic state. The spectrum was computed using the Maximum Entropy Method, especially good for short time series (cf. \citealp{oh}). The inset in the figure shows the corresponding phase portrait of the Lorenz system (a strange attractor). The dashed line indicates the best fit by the exponential Eq. (3) (deterministic chaos) with the fit's parameters $f_c = 506.6 (\pm 2.4) \times 10^{-5}$, and $\chi^2 =  4.877$.\\
  
  The random variations of the parameter $f_c$ in Eq. (3) result in the randomization of the deterministic chaos. A broadening of the deterministic chaos concept for {\it smooth} systems with randomly fluctuating  characteristic frequency  $f_c$ -- distributed chaos, yields stretched exponential spectra
\begin{equation}
E(f) \propto \exp-(f/f_{\beta}) ^{\beta}.  
\end{equation}  
In these systems, chaos is not deterministic, yet it remains smooth. The power-law (scaling) spectra characterize non-smooth systems (see, for instance,  Ref. \citealp{bs} where the Kolmogorov-like scaling spectrum has been analyzed for highly turbulent space plasma). 

  It should be noted that in a smooth forecast, the predictability horizon is expected to be prolonged indefinitely by decreasing the initial error. The potential for a smooth forecast of the solar magnetic field dynamics would arise from the stretched exponential spectral behavior. For dynamics characterized by power-law spectra, a smooth forecast is generally unachievable. Thus, it is essential to understand what type of spectra (stretched exponential or power-law) is characteristic of the dynamics of the solar magnetic field.\\
   
  A particular type of stretched exponential spectra with $\beta =1/2$ observed in the direct numerical simulations, laboratory experiments, and solar observations has been used in the present paper to validate that the considered MHD dynamo processes are dominated by the cross-helicity Eq. (1) (or its generalizations) in the framework of the Kolmogorov-like phenomenology.

\section{Cross-helicity and Kolmogorov's phenomenology}

\subsection{Scaling estimates}  

   An inertial range of scales is anticipated for high Reynolds numbers in hydrodynamic turbulence. In this range, the statistical characteristics of the motion are determined solely by the kinetic energy dissipation rate $\varepsilon$ (Kolmogorov's phenomenology) \citep{my}. In magnetohydrodynamics, a cross-helicity-inertial range of scales can be introduced. In this range, two parameters: the averaged cross-helicity $h_{cr} $ (or its modification $\tilde{h}_{cr}$) and the total energy dissipation rate $\varepsilon$ -- control the behavior of the magnetic field dynamics \citep{b2}. Using dimensional considerations, a scaling relationship between the characteristic value of magnetic field fluctuations $b_c$ and the characteristic frequency $f_c$ can be obtained
 \begin{equation}
 b_c \propto h_{cr} ~\varepsilon^{-1/2}~f_c^{1/2} 
 \end{equation}
where $\varepsilon$ is the total energy dissipation rate (for $\tilde{h}_{cr}$ the relationship is the same because $\tilde{h}_{cr}$ has the same dimensionality as $h_{cr}$).

\subsection{Spontaneous breaking of local reflectional symmetry}

   Even in the cases with zero (or negligible) net cross-helicity, spatially confined (localized) cross-helicity density $({\bf u}({\bf x},t)\cdot {\bf b}({\bf x},t))$ can be rather large in chaotic/turbulent flows \citep{mene,sch}. Since the velocity field ${\bf u}$ is a polar vector and the magnetic field ${\bf b}$ is an axial vector, the presence of nonzero local cross-helicity is linked to the absence of local reflection symmetry, indicating a form of local spontaneous reflection symmetry breaking.\\

   The localized spontaneous symmetry breaking leads to the formation of blobs with nonzero cross- and magnetic helicity \citep{moff1,moff2,mt,b1}. The magnetic surfaces of these blobs can be defined by the boundary conditions: ${\bf b_n}\cdot {\bf n}=0$, with ${\bf n}$ representing a unit vector perpendicular to the blob's boundary. The magnetic helicity (along with its adiabatic conservation) greatly lessens energy dissipation in these blobs \citep{moff1,moff2}, resulting in their extended lifespan within chaotic or turbulent settings.
   
     The localized in the blob with number $j$ and volume $V_j$  sign-defined cross-helicity 
\begin{equation}
H_{cr,j}^{\pm}  = \int_{V_j} {\bf u}({\bf x},t)\cdot {\bf b}({\bf x},t) d{\bf r}  
\end{equation}
is an ideal invariant \citep{mt} (here  `+'  or `-' denotes the blob's helicity sign). 

    Then, one can consider the total sign-defined ideal (adiabatic) invariant 
\begin{equation}
{\rm I^{\pm}} =  \frac{1}{V} \sum_j H_{cr,j}^{\pm}  
\end{equation}
 The summation takes into account the blobs with a specific sign only ('+' or '-'), and V denotes the total volume.
 
  The adiabatic invariant ${\rm I^{\pm}}$ defined by Eq. (8) can be used instead of the averaged magnetic helicity density $ h_{cr}$ in the estimate Eq. (6), for the case of the local reflection symmetry breaking
\begin{equation}
b_c \propto |{\rm I^{\pm}}| \varepsilon^{-1/2} f_c^{1/2}
\end{equation}
  Since the global (net) reflectional symmetry is still intact, ${\rm I^{+}} = - {\rm I^{-}}$.

\section{Distributed chaos in MHD}

    Intensification of the fluid's motion leads to random fluctuations of the characteristic frequency $f_c$ in Eq. (3). One must take this phenomenon into account. This can be achieved through ensemble averaging.
\begin{equation}
E(f) \propto \int_0^{\infty} \mathcal{P}(f_c) \exp -(f/f_c)df_c 
\end{equation}  

   Here, a probability {\it distribution} $\mathcal{P}(f_c)$ characterizes the random fluctuations of $f_c$. This is the reasoning for the term `distributed chaos'.

  In the case of the magnetic field dynamics governed by the cross-helicity, the relationship Eq. (6) (or Eq. (9)) can be utilized to ascertain the probability distribution $\mathcal{P}(f_c)$.\\

  The characteristic value of magnetic field $b_c$ can be taken half-normally distributed $\mathcal{P}(b_c) \propto \exp- (b_c^2/2\sigma^2)$ \citep{my}. It represents a normal distribution with a zero mean, which is truncated to have a nonzero probability density only for positive values of its variable. For example, if $b$ is a normally distributed variable, then the variable $b_c = |b|$ has a half-normal distribution \citep{jkb}. From Eq. (6) (or Eq. (9)), we obtain for this case
\begin{equation}
\mathcal{P}(f_c) \propto f_c^{-1/2} \exp-(f_c/4f_{\beta})  
\end{equation}
 It is the so-called chi-squared probability distribution function where $f_{\beta}$ denotes a new constant. \\

   Substituting Eq. (11) into Eq. (10), we obtain
\begin{equation}
E(f) \propto \exp-(f/f_{\beta})^{1/2}  
\end{equation}

%%%%%%%%%%%%%%% 2 %%%%%%%%%%%%%%%%%%
\begin{figure} \vspace{-3cm}\centering \hspace{-1.1cm}
\epsfig{width=.9\textwidth,file=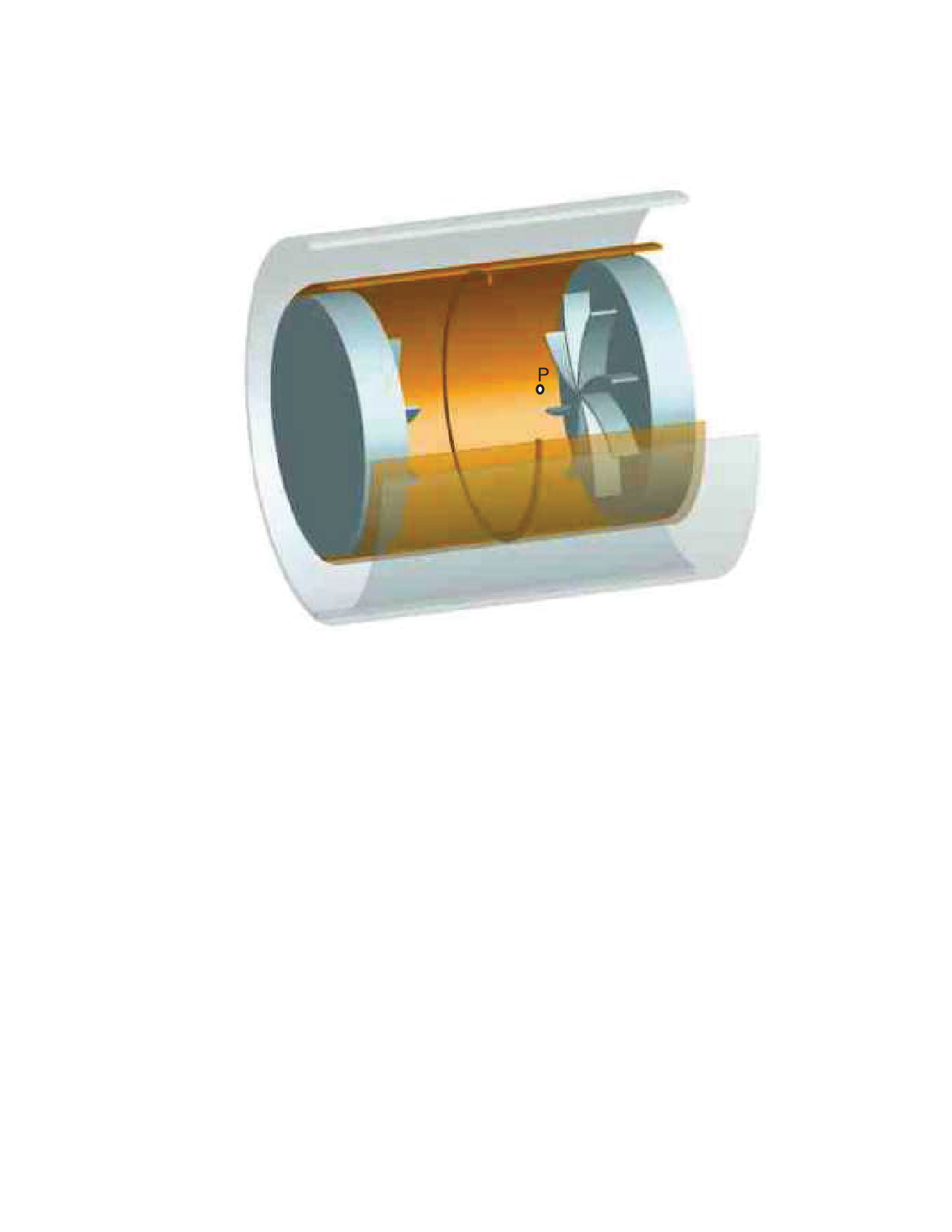} \vspace{-8.6cm}
\caption{Schematic representation of the setup's arrangement for experiments involving the von K\'{a}m\'{a}n flow.} 
\end{figure}
%%%%%%%%%%%%%%%%%%%%%%%%%%%%%%%%%%%

\section{Distributed chaos in MHD laboratory and numerical simulations} 

 In the paper \citep{min}, the results related to an MHD dynamo obtained in a laboratory experiment with a highly electrically conductive fluid (liquid sodium) were reported. A chaotic/turbulent flow was generated in the gap between two coaxial counter-rotating impellers (the so-called von K\'{a}rm\'{a}n flow; see a sketch of the experimental installation in Fig. 2).
 The magnetic Reynolds number $R_m \simeq 50$, attainable in this experiment, was sufficient to achieve a fully self-sustained chaotic MHD dynamo.\\ 
 
   Figure 3 shows (as the solid black curve) the power spectrum of the axial magnetic field fluctuations measured by the probe P (Fig. 2). The spectral data were taken from Fig. 8 of Ref. \citep{min}.  \\
 
%%%%%%%%%%%%%%% 3 %%%%%%%%%%%%%%%%%%
\begin{figure} \vspace{-2cm}\centering \hspace{-1.5cm}
\epsfig{width=.9\textwidth,file=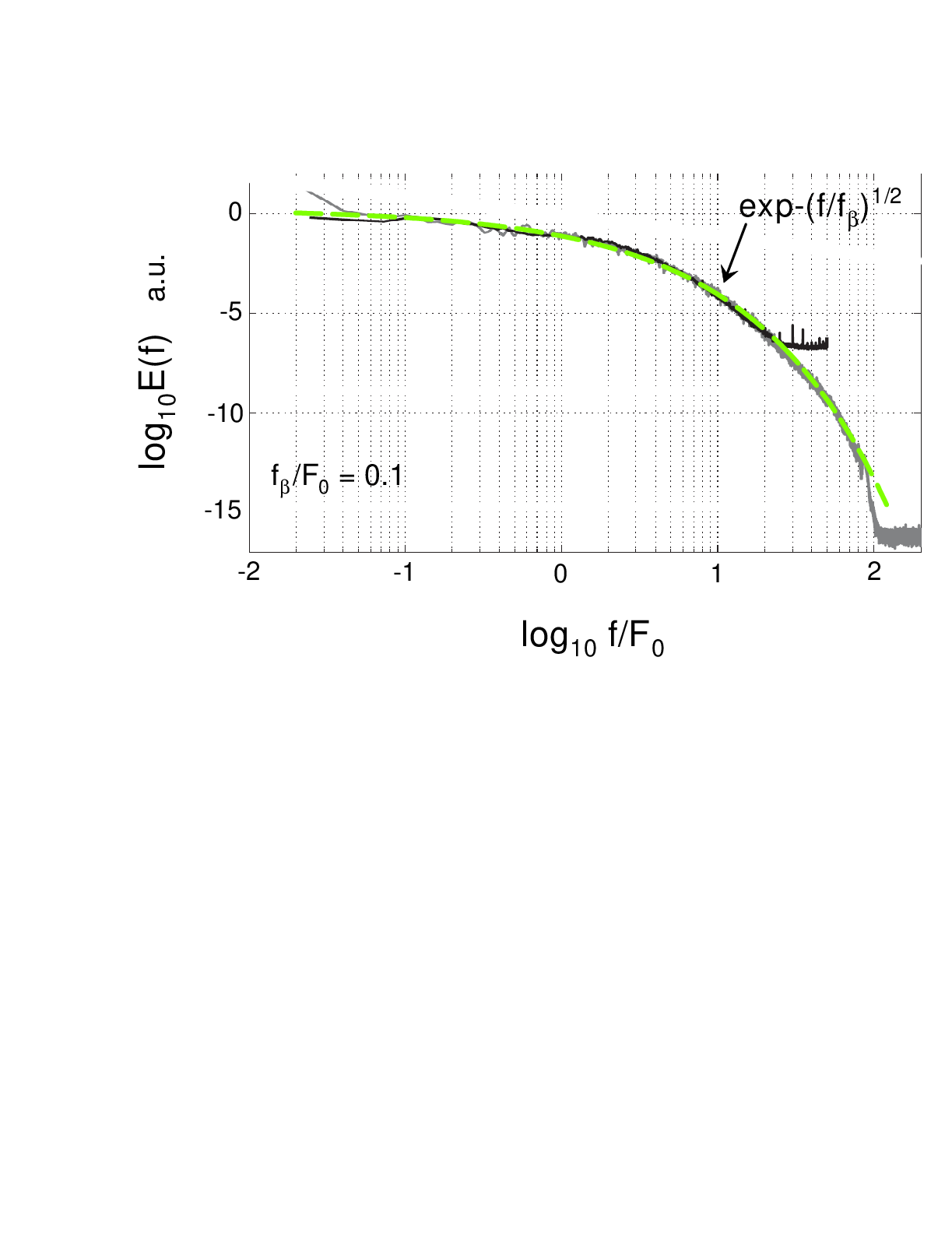} \vspace{-8.8cm}
\caption{Power spectrum (black curve) of the axial magnetic field component fluctuations measured in the bulk of the von K\'{a}rm\'{a}n flow in liquid sodium at $R_m \sim 50$ (MHD dynamo regime). The gray curve corresponds to the analogous power spectrum computed in a direct numerical simulation -- an analogue of the von K\'{a}rm\'{a}n flow.} 
\end{figure}
%%%%%%%%%%%%%%%%%%%%%%%%%%%%%%%%%%%  
 
    Additionally, the paper \citep{min} reports results of a direct numerical simulation that mimics the experiment. 
   The numerical simulation utilizes the standard MHD equations in the Alfvenic units for an incompressible ($\nabla \cdot {\bf u} = 0$, and
$\nabla \cdot {\bf b} = 0$) flow within a three-dimensional periodic domain with a side length of $2\pi$. 
The dimensionless equations are written as
\begin{equation}
\frac{\partial {\bf u}}{\partial t} + {\bf u} \cdot \nabla {\bf u} =
     -\nabla {\cal P} + {\bf j} \times {\bf b}+
      \nu \nabla^2 {\bf u} + {\bf F } ,
\end{equation}
\begin{equation}
\frac{\partial {\bf b}}{\partial t} + {\bf u} \cdot \nabla {\bf b} =
      {\bf b} \cdot \nabla {\bf b} +
      \eta \nabla^2 {\bf b} .
\label{eq:induction}
\end{equation}
Here, ${\cal P}$ is the  pressure normalized by the density, $\eta$ is the magnetic diffusivity, and $\nu$ is the dimensionless viscosity, 

The mechanical forcing generated in the experiment by the two impellers rotating in opposite directions (see Fig. 2) was replicated in the DNS using two Taylor-Green vortices.\\

   The same Fig. 3 additionally shows (as the solid gray curve) a power spectrum for the analogous signal obtained in the direct numerical simulation. The spectral data were taken from Fig. 8 of Ref. \citep{min}. The frequency in Fig. 3 was normalized by the `forcing' frequency $F_0 = u_{rms}/L$ ($u_{rms}$ is the root mean square of the velocity fluctuations, and 2$L$ is the side of the spatial domain) for the DNS and $F_0 =10$ Hz (the impellers' rotation rate) for the experiment.
   
  In this experiment and the corresponding direct numerical simulation, the velocity fluctuations are strong, and a well-defined mean velocity is absent in the bulk of the chaotic/turbulent flow. Therefore, Taylor's `frozen-in' hypothesis cannot be applied to these flows as well (see, for instance, Ref. \citealp{pl}). Hence, the spectra in Fig. 3 can be interpreted as true Euler frequency spectra. The dashed curve in Fig. 3 indicates the best fit by the stretched exponential spectrum Eq. (12) corresponding to cross-helicity-dominated distributed chaos. The best fit parameters are $f_{\beta}/F_0 = 0.093(\pm 0.002)$, and $\chi^2 = 0.038$.
  
   The boundary conditions and physical parameters in the experiment and direct numerical simulation are considerably different (moreover, the numerical MHD dynamo generates an equatorial dipole, whereas the experimental MHD dynamo generates an axial dipole). Nevertheless, the properly normalized spectra in Fig. 3 coincide extremely well. This can indicate a universal character of the spectrum Eq. (12).
  
     It is noteworthy to highlight that the von K\'{a}rm\'{a}n flow exhibits substantial (local) helicity and differential rotation, traits frequently observed in the interiors of stars and planets. For electrically conductive fluids and plasmas, these characteristics (under certain conditions) can significantly improve the conversion of kinetic energy from the fluid's (plasmas') motion into magnetic energy and support the MHD dynamo.\\
  
%%%%%%%%%%%%%%% 4 %%%%%%%%%%%%%%%%%%
\begin{figure} \vspace{-2.2cm}\centering \hspace{-1.3cm}
\epsfig{width=.8\textwidth,file=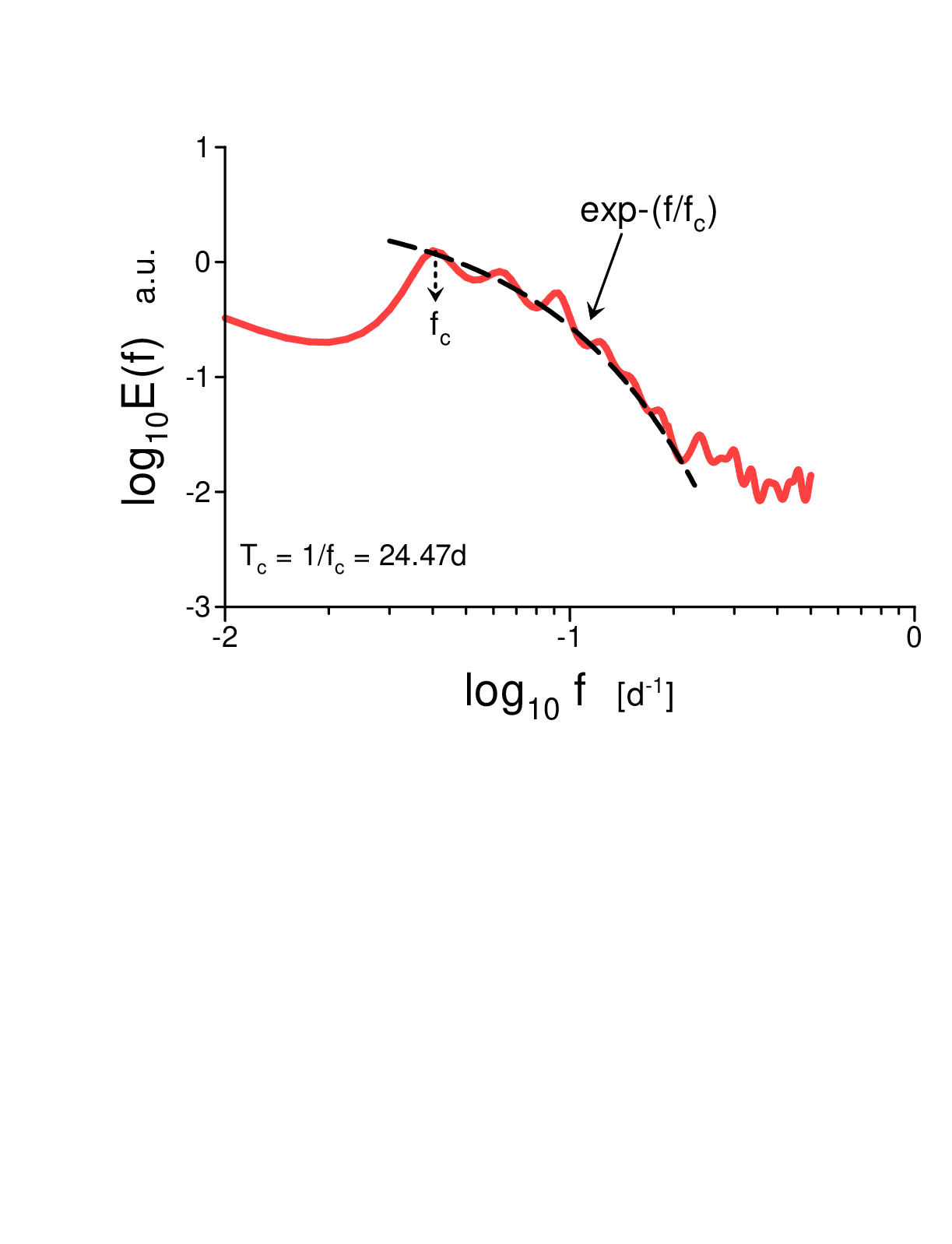} \vspace{-7cm}
\caption{Power spectrum of the full disk solar magnetic field  (the daily data for the period 2003--2017yy were taken from the site \citealp{sol}).} 
\end{figure}
%%%%%%%%%%%%%%%%%%%%%%%%%%%%%%%%%%%  

\section{Solar magnetic field}

\subsection{Direct measurements}

   Recent observations indicate that the global solar magnetic field has steadily decreased in strength over the past several solar cycles (see, for instance, recent papers \citealp{obr,murs,tan} and references therein). Therefore, one can expect indications of deterministic chaos in the recent solar dynamics (see Introduction). 
   
     Figure 4 shows the power spectrum of the full-disk solar magnetic field (the daily data for the period 2003--2017 were taken from the site \citealp{sol}). The mean net magnetic flux was measured with the Vector SpectroMagnetograph on the Synoptic Optical Long-term Investigations of the Sun (SOLIS) telescope. The power spectrum was computed using the Maximum Entropy Method, especially good for short time series \citep{oh}. 
     
     The dashed curve in Fig. 4 (the best fit) indicates the exponential spectrum Eq. (3) corresponding to the deterministic chaos (cf. Fig. 1). The dotted vertical arrow indicates the position of the $ f_c$ (the best fit parameters are $f_c = 408 (\pm 1) \times 10^{-4}~ d^{-1}$, and $\chi^2 = 0.348$). This value of the $f_c$ also corresponds to the main peak in the frequency spectrum and the period $T_c = 1/f_c = 24.47$d. This period is known as the sidereal period of the differential solar rotation at the solar equator. One can conclude that the coherent structures generated by the equatorial solar rotation determine the deterministic chaos in the global solar magnetic field. It should be noted that a confinement of chaos/turbulence to equatorial regions of the convective Sun’s outer envelope was recently discussed in the paper \citep{hhs}.  \\

%%%%%%%%%%%%%%% 5 %%%%%%%%%%%%%%%%%%
\begin{figure} \vspace{-2cm}\centering \hspace{-1.3cm}
\epsfig{width=.8\textwidth,file=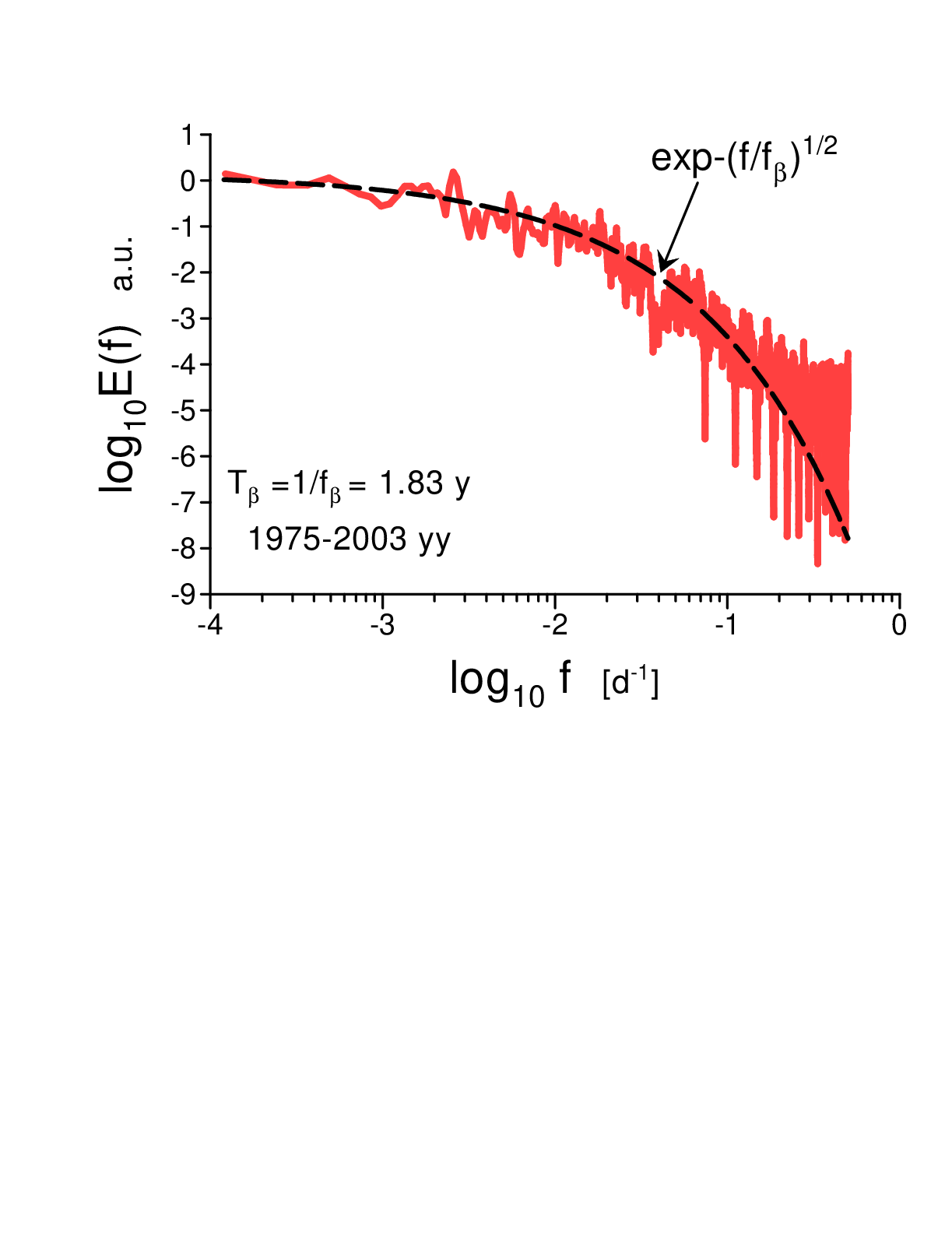} \vspace{-7.2cm}
\caption{Power spectrum of the full disk solar magnetic field  (the daily data for the period 1975--2003yy were taken from the site \citealp{mf}).} 
\end{figure}
%%%%%%%%%%%%%%%%%%%%%%%%%%%%%%%%%%% 

   Solar cycles 21 and 22 (1976-1996) exhibit a comparatively strong global magnetic field, especially when compared to the weaker fields observed in the above-considered cycles 23 and 24 (1996-2019). The daily global magnetic field data were collected during solar cycles 21 and 22 by the Wilcox Solar Observatory \citep{mf}. The analysis of this observational time series has two primary issues. The first issue is that around 19\% of the data points are uncertain. The second issue is the presence of notably large spectral peaks corresponding to the Carrington (an average synodic) period of solar rotation ($T \simeq 27$ days) and its harmonics. The latter issue indicates a significant contribution from sunspots since the Carrington period characterizes solar differential rotation at a latitude of 26 degrees, where the majority of sunspots are observed. A moving (running) average of the time series with the Carrington period can address both of these issues.\\
     
     Figure 5 shows the power spectrum of the full-disk solar magnetic field computed for the averaged time series (the daily data for the period 1975--1996 were taken from the site \citealp{mf}). The computation was performed using the Fast Fourier Transform (FFT) method (the time series is sufficiently long in this case). The dashed curve in Fig. 5 (the best fit) indicates the stretched exponential spectrum Eq. (12) corresponding to the distributed chaos dominated by cross-helicity.  The best fit parameters are $f_{\beta} = 150 (\pm 5) \times 10^{-5} ~d^{-1}$, and  $\chi^2 = 3.631$.
\\

\subsection{Reconstructions using sunspot number}

     Solar activity was relatively strong over the longer period from 1750 to 2000. Of course, there are no observational data for the global solar magnetic field for this period. However, one can consider existing sunspot number data characterizing the solar activity. 

     The magnetic field dynamics of the Sun is a primary driver of sunspot number variations, determining variability on time scales of months to decades. A magnetohydrodynamic dynamo process creates magnetic flux tubes within (or in the vicinity of) the Sun's convective zone. Subsequently, turbulent convection seizes the magnetic flux tubes and transports a portion of them from the convection zone surface into the photosphere, resulting in the creation of visible sunspots. The {\it internal dynamics} of sunspots can also be related to the small-scale near-surface dynamos (see, for instance, a recent review \citealp{remp} and references therein). The dynamics of {\it global} magnetic solar activity can be described by the time series of the sunspot number, which is a scalar quantity. However, one can expect that the degree of randomization $\beta$ of the global magnetic field dynamics is imprinted on the degree of randomization of the sunspot number dynamics for the time scales larger than the Carrington (rotational) period (see above). \\
 
  %%%%%%%%%%%%%%% 6 %%%%%%%%%%%%%%%%%%
\begin{figure} \vspace{-1.7cm}\centering \hspace{-1.3cm}
\epsfig{width=.8\textwidth,file=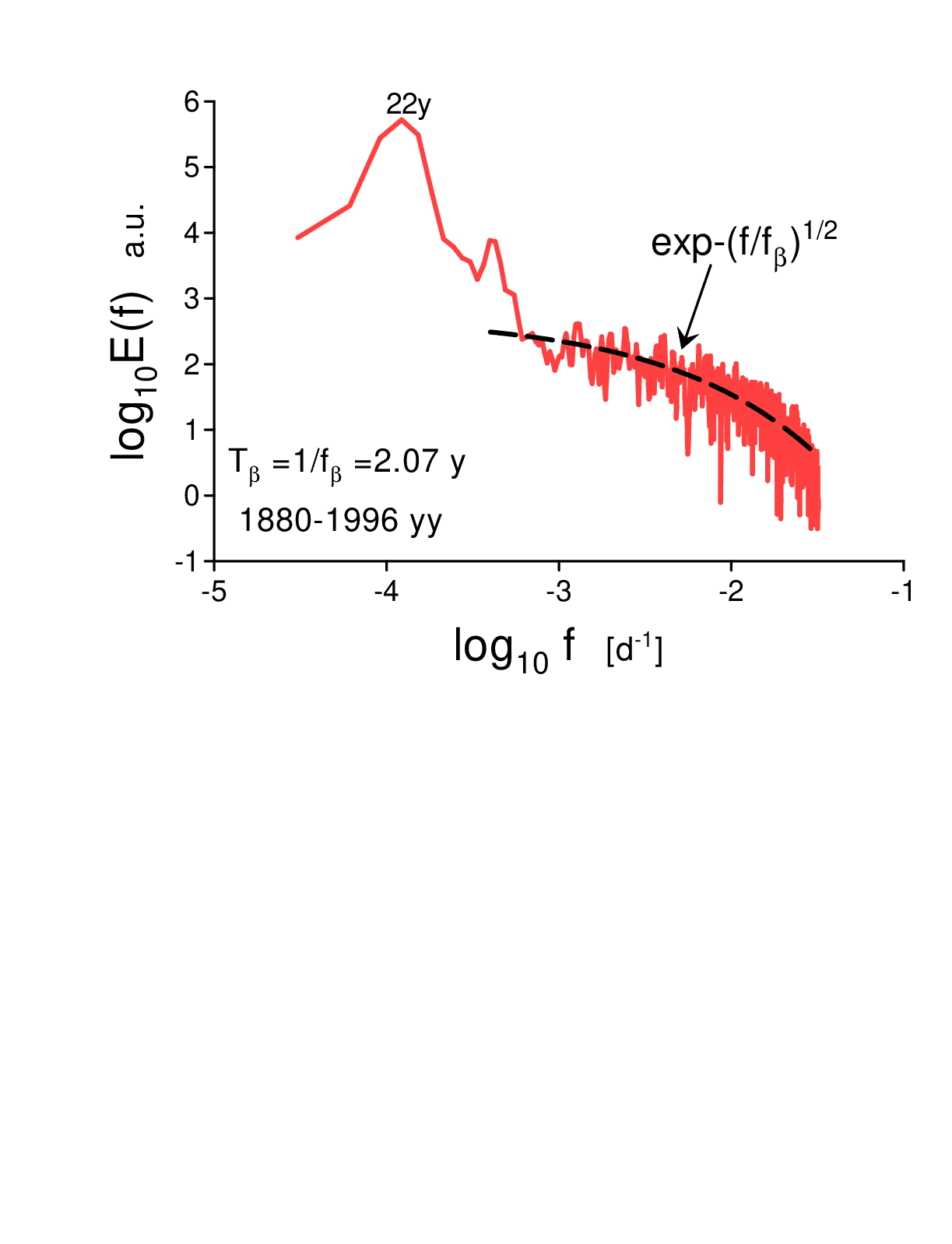} \vspace{-7.6cm}
\caption{Power spectrum of the magnetic field reconstructed from the daily sunspot number time series  (the daily data for the period 1880--2003yy were taken from the site \citealp{silso}).} 
\end{figure}
%%%%%%%%%%%%%%%%%%%%%%%%%%%%%%%%%%%   
     
     Several authors have argued that the sunspot number is proportional to the magnetic energy emerging in the photosphere (see e.g. \citealp{platt,tob,min2000,min2001}). Therefore, considering the sunspot number as quadratic in the magnetic field, the square root of the sunspot number series, with a sign change at each minimum \citep{brac,brac1985}, was taken in previous works to obtain a time series proportional to the spatially averaged magnetic field \citep{min2000,min2001}.
     
   Using this procedure (with the addition of the moving Carrington average before the sign-altering step), we converted the daily sunspot number time series presented on the site \citep{silso} for the period 1880--2003 into a proxy for the magnetic field.\\

%%%%%%%%%%%%%%% 7 %%%%%%%%%%%%%%%%%%
\begin{figure} \vspace{-0.8cm}\centering \hspace{-1.3cm}
\epsfig{width=.78\textwidth,file=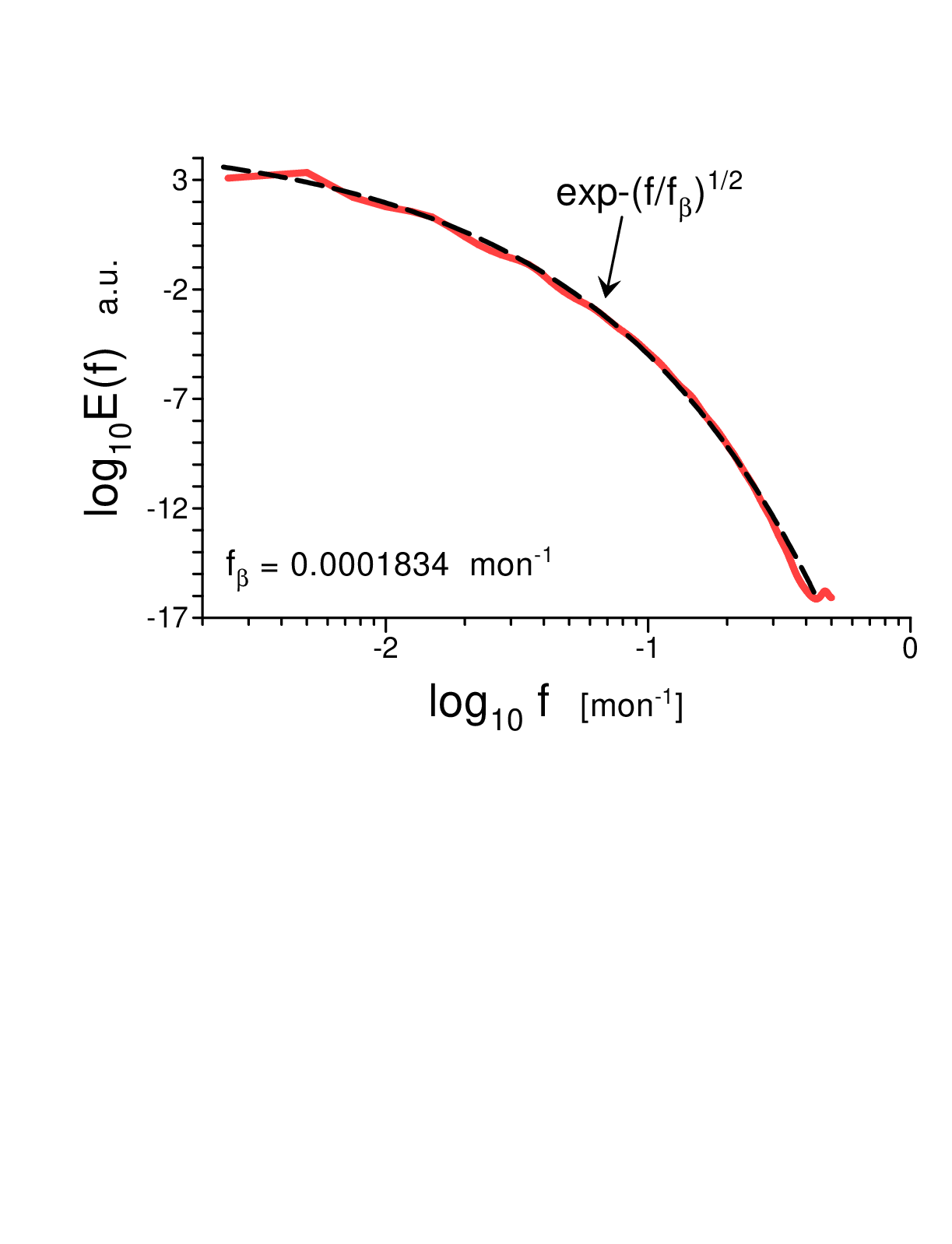} \vspace{-6.7cm}
\caption{Power spectrum of the reconstructed monthly time series of the covered (surrogate) solar global magnetic field.} 
\end{figure}
%%%%%%%%%%%%%%%%%%%%%%%%%%%%%%%%%%%  

     Figure 6 shows the power spectrum for the period 1880-- 2003 computed using the averaged time series with the Fast Fourier Transform (FFT) method (the time series is sufficiently long in this case). For such a long time series, a dominant fundamental peak at approximately 22 years and a secondary (small) mid-term peak localized in the 6–7 year band can be seen. The appearance of the secondary 6–7 year peak alongside the 22-year fundamental magnetic cycle is a known mathematical and physical feature of the signed Bracewell transformation. According to \citet{brac1985}, this specific power oscillation period does not represent an independent, secondary internal solar clock. Rather, it is the explicit manifestation of the third harmonic of the fundamental 22-year Hale cycle. 
     
     The dashed curve in Fig. 6 (the best fit) indicates the stretched exponential spectrum Eq. (12) corresponding to the distributed chaos dominated by cross-helicity for the time scales larger than the Carrington (rotational) period. The best fit parameters are $f_{\beta} = 133 (\pm 8) \times 10^{-5}~ d^{-1}$,
$\chi^2 = 0.652$.
 \\
     
    There is another approach to the problem. The underlying chaotic magnetohydrodynamics can be understood by reconstructing the associated multidimensional phase space. It has been estimated \citep{lamg,data} that an embedding dimension D=3 may be adequate for this purpose. The solar magnetic field cycle spans approximately 22 years, with 11 years during which the magnetic field polarity reverses. This indicates that the underlying magnetohydrodynamics should have the relevant symmetry group. Given that the sunspot number time series lacks this symmetry, one should find a {\it cover} system (having the symmetry group) that is dynamically (locally) equivalent to the system that lacks the symmetry group \citep{lg}. The phase portrait of the underlying cover (reconstructed) systems should be invariant under the inversion symmetry. Such a cover system was reconstructed in Ref. \citep{lamg} for the monthly time series of sunspots, spanning from 1750 to 2005. \\
  
   It should be noted that the covered magnetic field is a long-term construction implying a strong smoothing effect. Figure 7 shows the power spectrum corresponding to the reconstructed monthly time series of the covered (surrogate) magnetic field corresponding to the cover (reconstructed) system. The data were taken from the site \citep{data}. The spectrum was also computed using the Maximum Entropy Method.  The dashed curve in Fig. 7 (the best fit) indicates the stretched exponential spectrum Eq. (12) corresponding to the distributed chaos dominated by cross-helicity ($f_{\beta} = 183 (\pm 1) \times 10^{-6}$ ~mon$^{-1}$, $\chi^2 = 12.276$).

 \section{Conclusions}   
 
   The cross-helical chaotic/turbulent MHD dynamo is possible in the swirling flows (such as von K\'{a}rm\'{a}n's ones) of electrically conducting fluids (plasmas). This is a small-scale (fluctuation) dynamo dominated by the cross-helicity in a magneto-inertial range of scales. The chaotic dynamics of the magnetic field is smooth but not deterministic (distributed chaos). \\
   
   Analysis of the solar full-disc magnetic field for the last two solar cycles shows that these periods are characterised by weak magnetic activity and exhibit deterministic chaotic behavior concentrated around the equator. \\

  Analogous analysis of the solar cycles from 1975 to 1996, with a strong full-disc magnetic field, shows that these cycles exhibit distributed chaos dynamics dominated by the cross helicity. \\
   
   For longer periods, 1880-2003 and 1750-2005, the strong net magnetic field was reconstructed from the solar sunspot number time series using two distinct approaches (the former for daily and the latter for monthly time series). This analysis shows that the reconstructed strong magnetic field also exhibits the distributed chaos dominated by cross helicity. \\ 
   
 For solar cycles with a strong magnetic field, obtained through direct measurements (1975-2003) or by reconstruction from the sunspot time series (1880-2003), the distributed chaos characteristic period was approximately equal to 2 years. It can be compared to the well-known quasi-biennial periodicity, which has recently been related (using the data obtained by the Global Oscillation Network Group) to the two-year cycle of activity in the subsurface layers of the Sun \citep{sim}.\\

 The smooth forecast for the dynamics of the global solar magnetic field can be ensured by the stretched exponential spectra (see Introduction) observed over the past decades and centuries.


\begin{thebibliography}{99}
\bibitem[Atom(2025)]{data} Atom: 2025, Avilable at \url{http://www.atomosyd.net}

\bibitem[Bershadskii and Sreenivasan(2004)]{bs} Bershadskii, A., Sreenivasan, K. R.: 2004, Intermittency and the passive nature of the magnitude of the magnetic field. Phys. Rev. Lett. 93, 064501. 
\url{https://doi.org/10.1103/PhysRevLett.93.064501}

\bibitem[Bershadskii(2020)]{b2} Bershadskii, A.: 2020, Cross-helicity in Solar Active Regions. Res. Notes AAS 4, 10.
\url{https://doi.org/10.3847/2515-5172/ab6bd7}

\bibitem[Bershadskii(2025)]{b1} Bershadskii, A.: 2025, Emerging near-surface solar MHD dynamos. Foundations 5, 31. 
\url{https://doi.org/10.3390/foundations5030031}

\bibitem[Brandstater and Swinney(1987)]{swinney1} Brandstater, A., Swinney, H.L.: 1987, Strange attractors in weakly turbulent Couette-Taylor flow. Phys. Rev. A  35,  2207-2220. 
\url{https://doi.org/10.1103/PhysRevA.35.2207}.

\bibitem[Bracewell(1953)]{brac} Bracewell, R. N.: 1953, The sunspot number series. Nature 171, 649.
\url{https://doi.org/10.1038/171649a0}

\bibitem[Bracewell(1985)]{brac1985} Bracewell, R. N.: 1985, Sunspot number series envelope and phase. Aust. J. Phys. 38, 1009.
\url{https://doi.org/10.1071/PH851009}

\bibitem[Bhowmik et al.(2023)]{bho} Bhowmik P., Jiang J., Upton L, Lemerle A, Nandy D.: 2023. Physical models for solar cycle predictions, Space Science Reviews. 219, 40.
\url{https://doi.org/10.1007/s11214-023-00983-x}

\bibitem[Charbonneau and Sokoloff(2023)]{cs} Charbonneau, P., Sokoloff, D.: 2023, Evolution of solar and stellar dynamo theory. Space Sci. Rev. 219, 35.
\url{https://doi.org/10.1007/s11214-023-00980-0}

\bibitem[Hanasoge, Hotta, and Sreenivasan(2025)]{hhs} Hanasoge S., Hotta H., Sreenivasan K. R.: 2020, Turbulence in the Sun is suppressed on large scales and
confined to equatorial regions. Sci. Adv. 6, eaba9639. 
\url{https://doi.org/10.1126/sciadv.aba9639}

\bibitem[Johnson, Kotz, and Balakrishnan(1994)]{jkb} Johnson N. L., Kotz, S., Balakrishnan N.: 1994. Continuous Univariate Distributions. Vol. 1, Wiley NY
\bibitem[Letelliera and Gilmore(2000)]{lg} Letelliera C., and Gilmore R.: 2000, Covering dynamical systems: Twofold covers. Phys. Rev. E, 63, 16206.
\url{https://doi.org/10.1103/PhysRevE.63.016206}

\bibitem[Letellier et al (2006)]{lamg} Letellier C., Aguirre L. A., Maquet J., Gilmore R.: 2006, Evidence for low dimensional chaos in sunspot cycles. A\&A, 449, 379.
\url{https://doi.org/10.1051/0004-6361:20053947}

\bibitem[Lorenz(1963)]{lor} Lorenz, E.N.: 1963, Deterministic nonperiodic flow. J. Atm. Sci. 20, 130--141.

\bibitem[Maggs and Morales(2011)]{mm1} Maggs, J.E., Morales, G.J.: 2011, Generality of deterministic chaos, exponential spectra, and Lorentzian pulses in magnetically confined plasmas. Phys. Rev. Lett. 107, 185003. 
\url{https://doi.org/10.1103/PhysRevLett.107.185003}

\bibitem[Maggs and Morales(2012a)]{mm2} Maggs, J.E., Morales, G.J.: 2012a, Origin of Lorentzian pulses in deterministic chaos. Phys. Rev. E 86,  015401(R). 
\url{https://doi.org/10.1103/PhysRevE.86.015401}

\bibitem[Maggs and Morales(2012b)]{mm3} Maggs, J.E., Morales, G.J.: 2012b, Exponential power spectra, deterministic chaos and Lorentzian pulses in plasma edge dynamics. Plasma Phys. Control. Fusion 54, 124041.
\url{https://doi.org/10.1088/0741-3335/54/12/124041}

\bibitem[Meneguzzi et al.(1996)]{mene} Meneguzzi M., Politano H., Pouquet A., Zolver M.: 1996, A sparse-mode spectral method for the simulation of turbulent flows. J. Comp. Phys. 123,. 32.
\url{https://doi.org/10.1006/jcph.1996.0003}

\bibitem[Mininni, G\'omez, and Mindlin(2000)]{min2000} Mininni, P. D., G\'omez, D. O., and Mindlin, G. B.: 2000, Stochastic relaxation oscillator model for the solar cycle. Phys. Rev. Lett. 85, 5476.
\url{https://doi.org/10.1103/PhysRevLett.85.5476}

\bibitem[Mininni, G\'omez, and Mindlin(2001)]{min2001} Mininni, P. D., G\'omez, D. O., and Mindlin, G. B.: 2001, Simple model of a stochastically excited solar dynamo. Solar. Phys. 201, 203-223.
\url{https://doi.org/10.1023/A:1017515709106}

\bibitem[Mininni et al.(2014)]{min} Mininni P., Dmitruk P., Odier, P. Pinton J.-F.  Plihon N., Verhille, G., Volk R., Bourgoin M.: 2014, Long-term memory in experiments and numerical simulations of hydrodynamic and magnetohydrodynamic turbulence. Phys. Rev. E. 89, 053005.
\url{https://doi.org/10.1103/PhysRevE.89.053005}

\bibitem[Monin and Yaglom(2007)]{my} Monin, A.S., Yaglom, A.M.: 2007, Statistical Fluid Mechanics Vol. II: Mechanics of Turbulence. Dover Pub. NY.
\url{https://books.google.fr/books?id=6xPEAgAAQBAJ}

\bibitem[Moffatt(1969)]{moff1} Moffatt H. K.: 1969. The degree of knottedness of tangled vortex lines. J. Fluid Mech. 35, 117--129.
\url{https://doi.org/10.1017/S0022112069000991}

\bibitem[Moffatt(1985)]{moff2} Moffatt H. K.: 1985 Magnetostatic equilibria and analogous Euler flows of arbitrarily complex topology. Part 1. Fundamentals. J. Fluid Mech. 159, 359.
\url{https://doi.org/10.1017/S0022112085003251}

\bibitem[Moffatt and Tsinober(1992)]{mt}  Moffatt, H.K., Tsinober, A.: 1992, Helicity in laminar and turbulent flow. Annu. Rev. Fluid Mech. 24, 281-312.
\url{https://doi.org/10.1146/annurev.fl.24.010192.001433}

\bibitem[Mursula et al.(2024)]{murs} Mursula K., Pevtsov A. A., Asikainen T., Tahtinen I., Yeates A. R.: 2024, Transition to a weaker Sun: Changes in the solar atmosphere during the decay of the Modern Maximum. A\&A  685, A170.
\url{https://doi.org/10.1051/0004-6361/202449231}

\bibitem[Obridko et al.(2020)]{obr} Obridko, V.N., Sokoloff, D.D., Shelting, B.D. Shibalova, A.S., Livshits, I.M.: 2020, Cyclic variations in the main components of the solar large-scale magnetic field. MNRAS 492, 5582.
\url{https://doi.org/10.1093/mnras/staa147}

\bibitem[Obridko et al.(2025)]{obr2} Obridko, V., Shibalova, A., Sokoloff, D., Livshits, I.: 2025, Mean-field dynamo and forecasting of solar activity. JASTP 277, 106673.
\url{https://doi.org/10.1016/j.jastp.2025.106673}

\bibitem[Ohtomo et al.(1995)]{oh} Ohtomo N., Tokiwano K., Tanaka Y., et. al.: 1995, Exponential characteristics of power spectral densities caused by chaotic phenomena. J. Phys. Soc. Jpn., 64, 1104.
\url{https://doi.org/10.1143/JPSJ.64.1104}

\bibitem[Pinton and Labbe(1994)]{pl} Pinton J.-F., R. Labbe R.: 1994, Correction to the Taylor hypothesis in swirling flows. J. Phys. II 4, 1461.

\bibitem[Platt, Spiegel, and Tresser(1993)]{platt}  Platt, N., Spiegel, E. A., Tresser, C., 1993, The intermittent solar cycle. Geophys. Astrophys. Fluid Dyn. 73, 147-161. 
\url{https://doi.org/10.1080/03091929308203626}

\bibitem[Rempel et al.(2023)]{remp} Rempel M., Bhatia T., Bellot R., Rubio L. B., Korpi-Lagg M. J.: 2023, Small-scale dynamos: from idealized models to solar and stellar applications. Space Sci. Rev. 21, id. 36.
\url{https://doi.org/10.1007/s11214-023-00981-z}

\bibitem[Schekochihin(2022)]{sch} Schekochihin A. A.: 2022, MHD turbulence: a biased review. J. Plasma Phys. 88, 155880501.
\url{https://doi.org/10.1017/S0022377822000721}

\bibitem[Shebalin(2013)]{shebalin} Shebalin, J.V., 201,. Global invariants in ideal magnetohydrodynamic turbulence. Phys. Plasmas 20, 102305.
\url{https://doi.org/10.1063/1.482400}

\bibitem[SILSO(2025)]{silso} SILSO: 2025. Available at 
\href{https://www.sidc.be/SILSO/datafiles}{www.sidc.be}

\bibitem[Simoniello et al.(2013)]{sim} Simoniello R., Jain K., Tripathy C. Turck-Chi\'eze S., Baldner C., Finsterle W., Hill F., Roth M.: 2013, The quasi-biennial periodicity as a window on the solar magnetic dynamo configuration. ApJ 765, 100.
\url{https://doi.org/10.1088/0004-637X/765/2/100}

\bibitem[Solis(2025)]{sol} Solis: 2025, Available at 
\url{https://solis.nso.edu/pubkeep/v73/197001/kfv73700101/kfv73700101t000000_mf1.dat}

\bibitem[Stanford(2025)]{mf} Stanford: 2025, Available at 
\url{http://wso.stanford.edu/meanfld/MF_timeseries.txt}

\bibitem[Tanti et al.(2025)]{tan} Tanti H. A. Bisoi S. K., Datta A., Fujiki K.: 2025, Inner-heliospheric Signatures of Steadily Declining Solar Magnetic Fields and Their Possible Implications. ApJ 979, 198. 
\url{https://doi.org/10.3847/1538-4357/ada38a}

\bibitem[Tobias, Weiss, and Kirk(1995)]{tob} Tobias, S. M, Weiss, N. O., Kirk, V.: 1995, Chaotically modulated stellar dynamos. MNRAS 273, 1150-1166.  
\url{https://doi.org/10.1093/mnras/273.4.1150}

\bibitem[Yadav, Verma, and Wahi(2012)]{yvw} Yadav, R. K., Verma, M. K., Wahi P.: 2012, Bistability and chaos in the Taylor-Green dynamo. Phys. Rev. E 85, 036301.
\url{https://doi.org/10.1103/PhysRevE.85.036301}

\bibitem[Yokoi(2023)]{yok} Yokoi N.: 2023, Unappreciated cross‑helicity effects in plasma physics:
anti‑diffusion effects in dynamo and momentum transport. Review of Modern Plasma Physics 7, 33.
\url{https://doi.org/10.1007/s41614-023-00133-4}

\end{thebibliography}
\end{document}